# Finite-Element Analysis of MOF-Coated SAW Devices for Methane and Nitrous Oxide Sensing

Suman Acharya
*Precision Electrochemical Nanosensors*
*Tyndall National Institute*
Cork, Ireland
sacharya0921@gmail.com

Balasubramanian Srinivasan
*Precision Electrochemical Nanosensors*
*Tyndall National Institute*
Cork, Ireland
bala.srinivasan@tyndall.ie

Alan O'Riordan
*Precision Electrochemical Nanosensors*
*Tyndall National Institute*
Cork, Ireland
alan.oriordan@tyndall.ie

Veda Sandeep Nagaraja
*Strategic Development*
*Tyndall National Institute*
Cork, Ireland
veda.sandeep@tyndall.ie

*Abstract*— **A two-dimensional finite-element model was developed for MOF-coated 128-degree YX Lithium Niobate surface acoustic wave devices for methane and nitrous oxide sensing. ZIF-8 and UiO-66 coatings were modelled in COMSOL Multiphysics, with gas adsorption represented by a concentration-dependent increase in effective film density under a linear mass-loading approximation. Eigenfrequency analysis was used to identify the Rayleigh-dominant mode and evaluate coating-induced phase-velocity changes and gas-induced frequency shifts. A coating thickness of 150 nm was used on both the materials for comparison. The predicted sensitivities were 0.22 and 0.31 Hz/ppm for methane and nitrous oxide, respectively, with ZIF-8, compared with 0.65 and 3 Hz/ppm for UiO-66. The model isolates adsorption-induced mass loading while retaining the baseline mechanical properties of each MOF, enabling comparative assessment of MOF–gas combinations prior to experimental validation.**

## I. INTRODUCTION

SAW devices are attractive for gas sensing because acoustic displacement and associated electromechanical fields are concentrated near the surface of the piezoelectric substrate, making wave propagation sensitive to perturbations introduced by a functional coating. Gas adsorption can change the sensing layer's effective mass density, elastic and dielectric properties, and mechanical losses, thereby modifying acoustic velocity, resonance frequency, or attenuation. For sufficiently thin and electrically insulating coatings, mass loading provides a useful first-order basis for predicting gas-induced frequency shifts, although elastic effects may also contribute [1], [2]. MOFs offer high internal surface areas, tunable pore structures, and gas-adsorption capabilities [3]. Devkota *et al.* demonstrated MOF-coated $LiNbO_3$ acoustic sensors and highlighted the combined influence of gas uptake and sensing-film properties on their response [4]. Other finite-element studies have shown that sensing-layer thickness and elasticity can modify surface-wave velocity and must be considered alongside adsorbed mass [5]. COMSOL simulations of polymer-coated SAW devices have further demonstrated how concentration-dependent uptake can be represented by changes in effective sensing-layer density and analyzed through eigenfrequency calculations [6]. However, these considerations also highlight the need for comparisons of gas–MOF combinations under matched acoustic and geometric conditions, with the role of each assumed material property made explicit.

Previous device-level work considered simulation-guided optimization of $LiNbO_3$ SAW resonators and modified Butterworth–Van Dyke (mBVD) equivalent-circuit modelling [7]. The present study shifts attention to the gas-sensitive coating and its effect on a 128° YX $LiNbO_3$ SAW device. We develop a two-dimensional finite-element model in COMSOL Multiphysics to examine ZIF-8 and UiO-66 layers exposed to methane ($CH_4$) and nitrous oxide ($N_2O$). Both gases are environmentally relevant, and their different mass-uptake coefficients permit a comparative assessment of predicted sensor response. Adsorption is represented using a simple linear mass-uptake relationship that converts gas concentration into an increase in the effective sensing-layer density. The resulting frequency shift is evaluated by tracking the corresponding acoustic eigenmode, while all other film properties are held constant. This isolates the modelled mass-loading contribution rather than accounting for all possible adsorption-induced changes.

The study examines coating-thickness dependence and concentration-dependent frequency shifts for both MOFs and gases under identical acoustic conditions, providing a simple computational framework for screening gas–MOF combinations before fabrication.

## II. NUMERICAL MODELLING AND GAS-LOADING FORMULATION

### *A. Two-Dimensional Finite-Element Model*

The SAW device was modelled in COMSOL Multiphysics using a two-dimensional (2D) periodic unit cell comprising a 128° YX $LiNbO_3$ piezoelectric substrate, aluminum (Al) interdigital electrodes, and a continuous sensing layer deposited on the acoustic surface. The periodic unit cell was designed for an operating SAW frequency of 300 MHz, with the geometrical dimensions defined relative to the corresponding acoustic wavelength $\lambda$. The geometrical parameters of the 2D periodic SAW model are summarized in Table I. The width of the unit cell was defined as one acoustic wavelength, $\lambda$, with an electrode width of $\lambda/4$, corresponding to a metallization ratio of 0.5. The Al electrode thickness was fixed at 100 nm, while the substrate depth was extended to $5\lambda$. The computational geometry and applied boundary conditions are illustrated in Fig. 1.

The use of a single-wavelength unit cell reduces the computational domain while representing the periodically repeated SAW structure through periodic boundary conditions. A substrate depth of $5\lambda$ was selected to minimize the influence of the lower boundary on the surface acoustic mode. This choice is appropriate for investigating Rayleigh-type surface waves, whose displacement amplitudes are predominantly concentrated near the piezoelectric surface and decay with increasing substrate depth, consistent with the finite-element modelling approach reported by Devkota et al. [5]. The periodic formulation therefore provides a computationally efficient representation of the local wave propagation characteristics without explicitly modelling the complete interdigital transducer.

*TABLE I. GEOMETRIC PARAMETERS OF THE 2D SAW MODEL.*

| Parameter | Symbol | Dimension |
|---|---|---|
| Operating frequency | $f_0$ | 300 MHz |
| Acoustic wavelength | l | 13.06 mm |
| IDT width | l/4 | 3.26 mm |
| IDT height | h | 100 nm |
| Electrode gap | l/4 | 3.26 mm |
| Substrate depth | 5l | 65.3 mm |
| MOF thickness | $M_t$ | 150 nm |

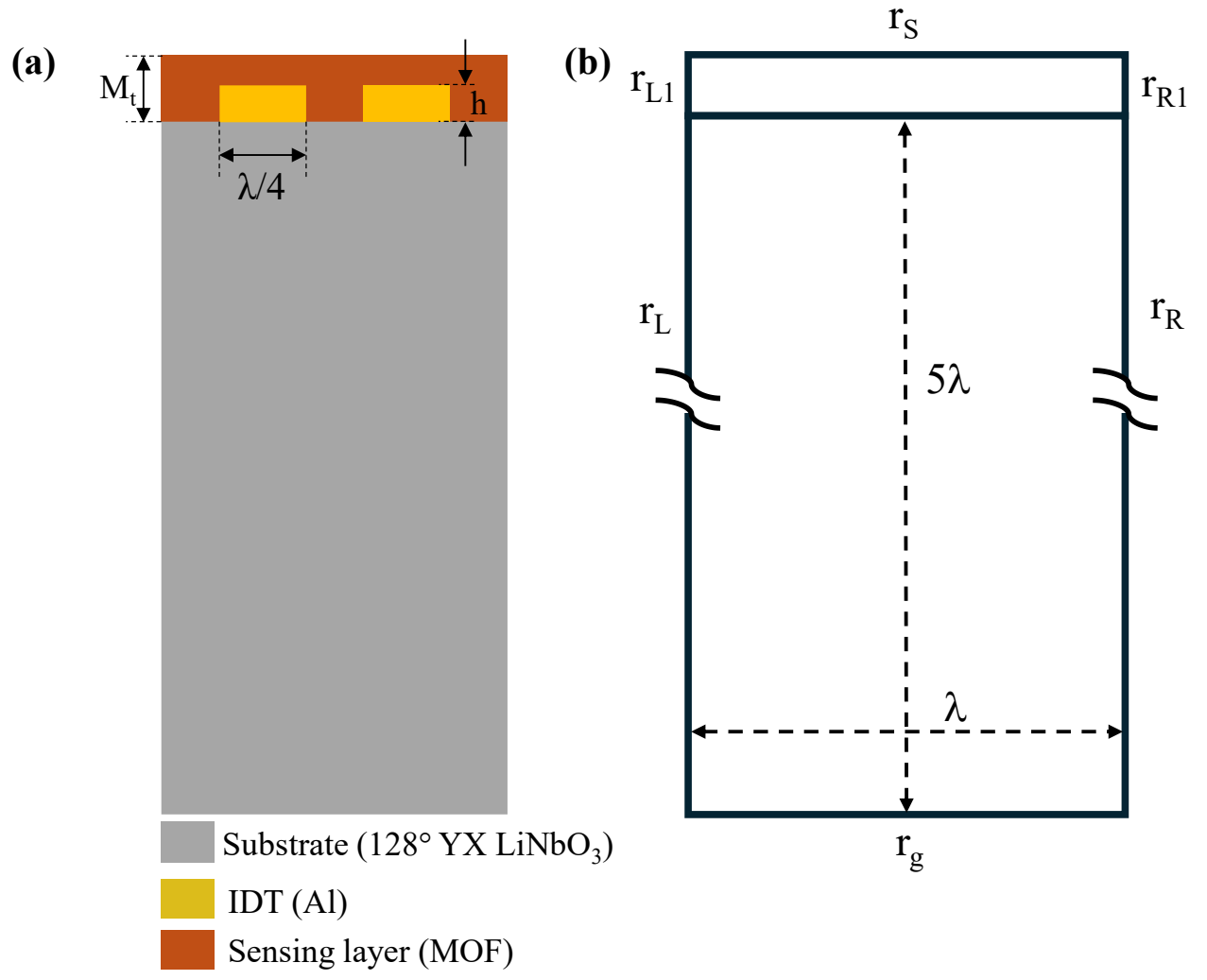

Fig. 1. (a) Geometry considered for two-dimensional finite-element modelling of the 128° YX $LiNbO_3$ SAW device and (b) mechanical and electrical boundary conditions applied to the periodic unit cell.

The coupled electromechanical response was solved using the Solid Mechanics and Electrostatics interfaces with piezoelectric coupling. The anisotropic elastic stiffness, piezoelectric, and dielectric properties of $LiNbO_3$ were manually defined in COMSOL Multiphysics to represent the 128° YX crystal orientation. Periodic conditions were imposed on the two lateral boundaries along the direction of wave propagation, as illustrated in Fig. 1(b). The upper surface was mechanically free, whereas the lower substrate boundary was mechanically fixed. For the electrical boundary configuration, a potential of 1 V was applied on the drive electrode, while the adjacent electrode was grounded at 0 V. The mechanical and electrical boundary conditions employed in the model are summarized in Table II.

A frequency domain study was performed to determine the acoustic modes supported by the periodic structure. The mode considered for the sensing analysis was identified as a Rayleigh-dominant SAW mode from its displacement distribution and strong confinement near the substrate surface. The same mode was tracked following deposition of the MOF sensing layer and subsequent adsorption-induced mass loading. For each gas concentration, the corresponding effective MOF density was updated, and the eigenfrequency problem was independently re-solved using identical geometry, mesh, boundary conditions, and solver settings. Consistent mode tracking was maintained because the gas-induced frequency shifts are small relative to the operating frequency. Consequently, each point in the simulated gas-response curves represents a separate FEM eigenfrequency solution for a specific gas-loading condition. This procedure enables the acoustic response to surface mass loading to be evaluated without explicitly modelling the complete finite transducer or its driven frequency response.

*TABLE II. MECHANICAL AND ELECTRICAL BOUNDARY CONDITIONS USED IN THE 2D FEM MODEL*

| Boundary | Mechanical | Electrical |
|---|---|---|
| $r_L$, $r_{L1}$, $r_R$, $r_{R1}$ | Periodic | Periodic |
| $r_s$ | Free | Zero charge |
| $r_g$ | Fixed | Zero charge |

### *B. MOF Sensing Layers and Gas loading*

ZIF-8 and UiO-66 were modelled as homogeneous, linearly elastic, and isotropic sensing layers. For a thin sensing film, where the film thickness h is much smaller than the SAW wavelength (h ≪ λ), the relative change in SAW velocity caused by deposition of the sensing layer can be approximated as [8]

$$\frac{\Delta v}{v} = -c_m f_0 h \rho_0 + c_e f_0 h \left[\frac{4G}{v_0^2}\left(\frac{3K+G}{3K+4G}\right)\right] \tag{1}$$

where $f_0$ and $v_0$ are the operating frequency and SAW velocity, respectively; $c_m$ and $c_e$ are the mass and elasticity sensitivity coefficients of the substrate; and $r_0$, K, and G are the density, bulk modulus, and shear modulus of the sensing layer, respectively. The first term represents the mass-loading contribution, whereas the second accounts for the elastic contribution of the sensing film. The expression illustrates that the acoustic response depends not only on the additional coating mass but also on its mechanical properties. The material properties assigned to ZIF-8 and UiO-66 in the FEM model are summarized in Table III.

*TABLE III. MATERIAL PROPERTIES ASSIGNED TO THE MOF SENSING LAYERS.*

| **MOF** | **Density ($kg/m^3$)** | **Young's Modulus (GPa)** | **Poisson's ratio** | **Relative Permittivity** |
|---|---|---|---|---|
| ZIF-8 | 950 | 3.5 | 0.3 | 2.7 |
| UiO-66 | 1240 | 37.10 | 0.35 | 1.85 |

When a sensing layer adsorbs gas molecules, its effective mass increases and introduces additional loading on the SAW surface. Gas uptake in porous materials can be determined experimentally through adsorption isotherms or estimated using molecular simulation methods such as grand canonical Monte Carlo (GCMC) [9]. The simulations were conducted over a concentration range of 0–300 ppm, within which adsorption was assumed to remain in the low-pressure Henry-law regime, such that gas uptake varies linearly with gas partial pressure and, at constant total pressure, with concentration [10]. In the present model, this behaviour was represented using a gas–MOF mass-uptake coefficient, $K_m$, which relates gas concentration to the corresponding increase in sensing-layer density:

$$\Delta \rho = K_m C \tag{2}$$

where $\Delta\rho$ is the loaded mass density, $K_m$ is the mass of gas adsorbed per unit volume of MOF per unit gas concentration, and C is the gas concentration in vol-%. The coefficients $K_m$ are expressed in (mg/cm$^3$)/vol-%. Gas concentrations were converted into units consistent with $K_m$ before evaluating the density perturbation. Specifically, 1 vol-% corresponds to 10,000 ppm, and a numerical value of 1 mg/cm$^3$ is equivalent to 1 kg/m$^3$. This conversion ensures consistency between the adsorption parameters and the SI material properties assigned in COMSOL.

Gas adsorption may also modify the elastic properties of the sensing film through adsorption-induced deformation or swelling [11]. In the present study, these gas-induced changes were neglected, and the incremental acoustic response was assumed to be governed by mass loading. Accordingly, the effective density assigned to the sensing-layer domain in COMSOL was $\rho_{eff} = \rho_0 + \Delta\rho$, while the remaining elastic and dielectric properties were kept unchanged. The baseline mechanical properties of the two MOFs were nevertheless retained in the model so that the coating-dependent acoustic response could be evaluated under their respective material conditions. The gas–MOF mass-uptake coefficients used as FEM inputs are summarized in Table IV.

*TABLE IV. GAS–MOF MASS-UPTAKE COEFFICIENTS, $K_m$ IN (mg/cm$^3$)/vol-% USED IN THE FEM SIMULATIONS.*

| **MOF** | **$CH_4$** | **$N_2O$** |
|---|---|---|
| ZIF-8 | 0.015 | 0.557 |
| UiO-66 | 0.043 | 2.308 |

The gas-induced resonance-frequency shift was determined relative to the corresponding gas-free MOF-coated device as

$$\Delta f = f_0 - f_g \tag{3}$$

where $f_0$ is the resonance frequency of the gas-free MOF-coated device and $f_g$ is the resonance frequency after incorporating the adsorption-induced mass loading. The gas-free coated state was used independently as the reference for each MOF, thereby separating the frequency shift caused by initial coating deposition from the additional shift associated with gas uptake. The gas sensitivity was calculated from the slope of the resonance-frequency shift with respect to gas concentration and is reported in Hz/ppm. This procedure enables comparison of the predicted gas responses of different MOF layers under identical acoustic and geometric conditions.

## III. RESULTS AND DISCUSSION

The simulated structure supports a strongly surface-confined acoustic mode, with the displacement concentrated near the interdigital electrode region and decreasing rapidly with depth into the $LiNbO_3$ substrate, as shown in Fig. 2(a). This distribution is consistent with the expected surface localization of Rayleigh-type SAW propagation and supports the suitability of the selected eigenmode for evaluating surface perturbations. Such confinement is particularly relevant for gas sensing because mechanical loading introduced by functional coating directly modifies the region in which a significant fraction of the acoustic energy is concentrated.

The influence of sensing-layer thickness was subsequently examined using ZIF-8. As shown in Fig. 2(b), the surface acoustic wave velocity decreases progressively with increasing ZIF-8 thickness. This trend reflects the modification of the acoustic boundary conditions by the coating, including its additional inertial loading and elastic properties. As the coating becomes thicker, a greater amount of material interacts mechanically with the surface wave, leading to a progressively modified propagation velocity. Similar considerations regarding the contributions of coating mass and viscoelastic properties to SAW propagation have been discussed previously [8]. A sensing-layer thickness of 150 nm was selected for the subsequent gas-sensing simulations and maintained for both ZIF-8 and UiO-66 to enable comparison under identical geometric conditions. This thickness represents a common reference configuration rather than an independently established optimum for gas sensitivity. At a sensing-layer thickness of 150 nm, the calculated gas-free resonance frequencies were 295.84 MHz for ZIF-8 and 296.92 MHz for UiO-66. Although the coating thickness was identical, the different baseline frequencies reflect the combined influence of the assigned material properties on the acoustic response. In particular, the coating density and elastic stiffness affect the mechanical loading and effective wave propagation characteristics of the structure. Consequently, the resonance frequency of a coated SAW device is determined not only by the coating thickness but also by its mechanical properties particularly in the case of UiO-66 coated device. These gas-free frequencies were therefore used independently as reference values for evaluating the additional adsorption-induced frequency shifts.

Figures 2(c) and 2(d) show the simulated responses to increasing $CH_4$ and $N_2O$ concentrations for the ZIF-8- and UiO-66-coated devices, respectively. In all cases, the magnitude of the resonance-frequency shift increases approximately linearly with gas concentration. This behaviour is consistent with the adopted low-concentration mass-loading formulation, in which the adsorption-induced increase in effective sensing-layer density is proportional to gas concentration. Under this assumption, the additional gas mass perturbs the acoustic eigenfrequency while the elastic and dielectric properties of the coating remain unchanged. The approximately linear response therefore characterizes the behaviour of the adopted FEM formulation over the investigated concentration range, rather than independently establishing linear adsorption under experimental conditions. The gas sensitivity was obtained from the slope of the calculated resonance-frequency shift with concentration. The extracted sensitivities for both MOF coatings are summarized in Table V.

TABLE V. SIMULATED FREQUENCY SENSITIVITIES OF THE MOF-COATED SAW DEVICES.

| **MOF** | **$CH_4$ (Hz/ppm)** | **$N_2O$ (Hz/ppm)** |
|---|---|---|
| ZIF-8 | 0.22 | 0.31 |
| UiO-66 | 0.65 | 3.0 |

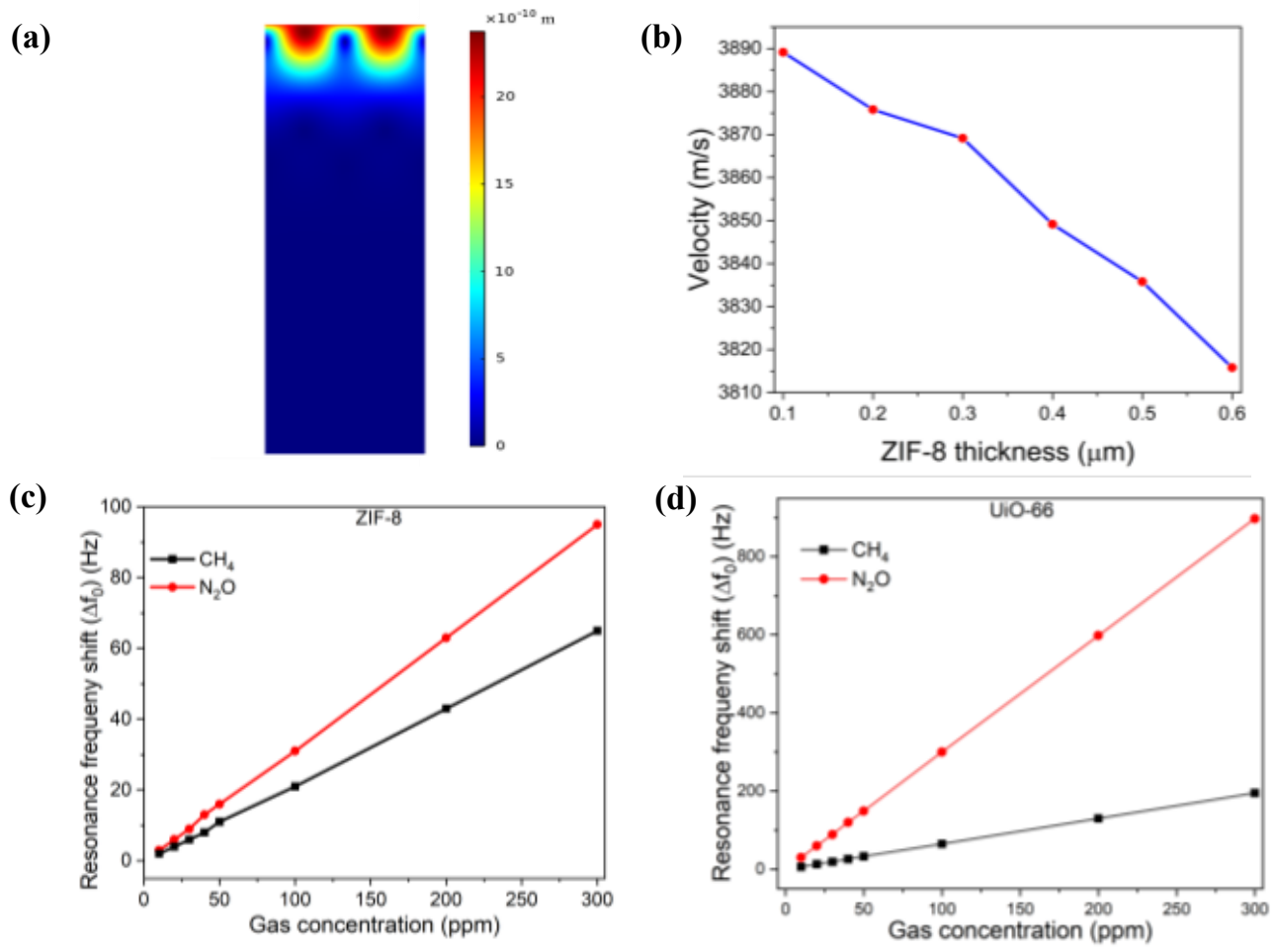

Fig. 2. (a) Simulated displacement magnitude of the surface-confined acoustic mode in the 128° YX $LiNbO_3$ structure; (b) variation of surface acoustic wave velocity with ZIF-8 sensing-layer thickness; and simulated resonance-frequency shift magnitude as a function of $CH_4$ and $N_2O$ concentration for (c) ZIF-8- and (d) UiO-66-coated SAW devices.

The UiO-66-coated device exhibited greater simulated frequency sensitivity to both gases, yielding 0.65 Hz/ppm for $CH_4$ and 3.00 Hz/ppm for $N_2O$, compared with 0.22 Hz/ppm and 0.31 Hz/ppm, respectively, for ZIF-8. The corresponding UiO-66 sensitivity was approximately 3.0 times higher for $CH_4$ and 9.7 times higher for $N_2O$. The highest predicted response was obtained for the UiO-66/$N_2O$ combination. These differences demonstrate that the frequency response depends on the particular gas–MOF combination and the acoustic properties of the coated structure. Within the present FEM framework, gas exposure is represented through adsorption-induced density changes, while the baseline mechanical properties remain specific to each MOF. Accordingly, the reported sensitivities characterize the response of the complete coated acoustic structure under the prescribed model conditions and should not be interpreted as direct measures of intrinsic adsorption capacity or experimentally demonstrated gas-detection performance.

The present FEM framework isolates adsorption-induced mass loading and does not account for associated changes in the elastic, dielectric, or damping properties of the sensing layer, although these mechanisms may also perturb SAW propagation [8], [11]. Adsorption/desorption kinetics, temperature, humidity, and competitive adsorption are likewise excluded. These effects may modify both the magnitude and temporal response of the sensor and should therefore be addressed experimentally. Within these assumptions, the model provides a computationally efficient basis for comparing MOF coatings and gas-loading conditions prior to fabrication.

## IV. SUMMARY

A two-dimensional finite-element model of 128° YX $LiNbO_3$ SAW devices coated with ZIF-8 and UiO-66 was developed for $CH_4$ and $N_2O$ sensing. The results show that coating properties and adsorption-induced mass loading govern the acoustic response, with UiO-66 exhibiting the highest predicted sensitivity of 3.00 Hz/ppm to $N_2O$. The model provides a computationally efficient framework for comparing MOF sensing layers prior to experimental validation.

## ACKNOWLEDGMENT

This publication has emanated in part from research supported by Science Foundation Ireland (SFI) under Grant No. 13/RC/2077_P2 (CONNECT), co-funded by the European Regional Development Fund, and Grant No. 21/RC/10303_P2 (VistaMilk), co-funded by the Department of Agriculture, Food and the Marine on behalf of the Government of Ireland.

## REFERENCES

[1] D. Mandal and S. Banerjee, "Surface acoustic wave (SAW) sensors: Physics, materials, and applications," *Sensors*, vol. 22, no. 3, Art. no. 820, 2022, doi: 10.3390/s22030820.

[2] S. Acharya, B. Srinivasan, D. Shanahan, et al., "Surface acoustic wave gas sensors: Innovations in functional materials, sensing dynamics, and signal analysis," *arXiv*, arXiv:2510.04940, 2025, doi: 10.48550/arXiv.2510.04940.

[3] J. F. Olorunyomi, S. T. Geh, R. A. Caruso, and C. M. Doherty, "Metal–organic frameworks for chemical sensing devices," *Mater. Horiz.*, vol. 8, no. 9, pp. 2387–2419, 2021, doi: 10.1039/D1MH00609F.

[4] J. Devkota, K.-J. Kim, P. R. Ohodnicki, J. T. Culp, D. W. Greve, and J. W. Lekse, "Zeolitic imidazolate framework-coated acoustic sensors for room temperature detection of carbon dioxide and methane," Nanoscale, vol. 10, pp. 8075–8087, 2018, doi: 10.1039/C7NR09536H.

[5] J. Devkota, P. R. Ohodnicki, J. A. Gustafson, C. E. Wilmer, and D. W. Greve, "Designing a SAW sensor array with MOF sensing layers for carbon dioxide and methane," in *Proc. Joint Conf. IEEE Int. Frequency Control Symp. and European Frequency and Time Forum (EFTF/IFC)*, 2019, pp. 1–4, doi: 10.1109/FCS.2019.8856088.

[6] C. Bostan, B. Serban, V. Avramescu, and I. Georgescu, "Simulation of SAW gas sensors with polymer layers using the finite element method," in *Proc. Int. Semiconductor Conf. (CAS)*, Sinaia, Romania, 2010, pp. 503–506, doi: 10.1109/SMICND.2010.5650470.

[7] S. Acharya, D. Shanahan, B. Srinivasan, U. Roedig, A. O'Riordan, and V. S. Nagaraja, "Toward sustainable and intelligent SAW gas sensors: Simulation-guided design and machine learning-assisted Butterworth–Van Dyke modeling," poster presented at the E-MRS 2025 Fall Meeting & Exhibit, Warsaw, Poland, Sep. 16, 2025, Poster 35_711.

[8] S. J. Martin, G. C. Frye, and S. D. Senturia, "Dynamics and response of polymer-coated surface acoustic wave devices: Effect of viscoelastic properties and film resonance," *Anal. Chem.*, vol. 66, no. 14, pp. 2201–2219, 1994, doi: 10.1021/ac00086a003.

[9] T. R. Zeitler, M. D. Allendorf, and J. A. Greathouse, "Grand canonical Monte Carlo simulation of low-pressure methane adsorption in nanoporous framework materials for sensing applications," *J. Phys. Chem. C*, vol. 116, no. 5, pp. 3492–3502, 2012, doi: 10.1021/jp208596e.

[10] J. A. Gustafson and C. E. Wilmer, "Optimizing information content in MOF sensor arrays for analyzing methane-air mixtures," *Sens. Actuators B Chem.*, vol. 267, pp. 483–493, 2018, doi: 10.1016/j.snb.2018.04.049.

[11] J. W. Grate, M. Klusty, R. A. McGill, M. H. Abraham, G. Whiting, and J. Andonian-Haftvan, "The predominant role of swelling-induced modulus changes of the sorbent phase in determining the responses of polymer-coated surface acoustic wave vapor sensors," *Anal. Chem.*, vol. 64, no. 6, pp. 610–624, 1992, doi: 10.1021/ac00030a009.